\documentclass[amstex,epsf,amssymb,usenatbib]{mn2e}

\usepackage{epsfig}
\usepackage{epsf}
\usepackage{graphicx}
\usepackage{tabularx}
\usepackage{amsmath}
\usepackage{amssymb}
\usepackage{amstext}
\usepackage{amsfonts}
\usepackage{comment}
\usepackage{color,soul}

\title{Power spectra of outflow-driven turbulence}

\author[A. Moraghan, Jongsoo Kim \& Suk-Jin Yoon]
{Anthony Moraghan $^{1,2,3}$
Jongsoo Kim $^{3,4}$
\& Suk-Jin Yoon $^{2}$\thanks{E-mail: sjyoon@galaxy.yonsei.ac.kr},\\
$^1$Academia Sinica Institute of Astronomy and Astrophysics, P.O. Box 23-141, Taipei 106, Taiwan\\
$^2$Center for Galaxy Evolution Research and Department of Astronomy, Yonsei University, Seoul 120-749, Republic of Korea\\
$^3$Korea Astronomy and Space Science Institute, Daejeon 305-348, Republic of Korea\\ 
$^4$Astronomy and Space Science Major, Korea University of Science and Technology, Daejeon 305-350, Republic of Korea\\
}

\date{Accepted 2015 March 24. Received 2015 March 23; in original form 2015 February 13}

\pagerange{\pageref{firstpage}--\pageref{lastpage}}
\pubyear{2015}

\begin{document}

\maketitle

\label{firstpage}

\begin{abstract}
We investigate the power spectra of outflow-driven turbulence through 
high-resolution three-dimensional isothermal numerical simulations where 
the turbulence is driven locally in real-space by a simple spherical outflow model. The resulting 
turbulent flow saturates at an average Mach number of  $\sim 2.5$ and is analysed through density 
and velocity power spectra, 
including an investigation of the evolution of the solenoidal and compressional components.
We obtain a shallow density power spectrum with a slope of $\sim -1.2$ attributed to the presence of a network of 
localised dense filamentary structures formed by strong shock interactions.
The total velocity power spectrum slope is found to be $\sim -2.0$, representative of 
Burgers shock dominated turbulence model.
The density weighted velocity power spectrum slope is measured as $\sim -1.6$, slightly less than the 
expected Kolmogorov scaling value (slope of -5/3) found in previous works.
The discrepancy may be caused by the nature of our real space driving model 
and we suggest there is no universal scaling law for supersonic compressible turbulence.
We find that on average, solenoidal modes slightly dominate in our turbulence model as the interaction 
between strong curved compressible shocks generates solenoidal modes, and compressible modes 
decay faster. 

\end{abstract}


\section{Introduction}              

Observations, theory, and simulations have shown that the physics of star forming regions are 
dominated by supersonic turbulence.
The turbulence supports against gravitational collapse of the molecular cloud and leads to a hierarchical structure 
of diffuse regions, filaments, and clumps, as kinetic energy is distributed from kilo-parsec scales 
to sub-Astronomical Unit (AU) scales\,\citep{2013ApJ...771..123B}.
Simulations have shown supersonic turbulence to decay rapidly, in less than a crossing time, and so the 
turbulence must be continuously replenished in order to account for the relatively low star formation rates (SFR) 
that are observed in molecular clouds \citep{2004RvMP...76..125M}.

Protostellar jets and outflows, being synonymous with star formation, are one possible candidate
as a mechanism to drive and sustain supersonic turbulence within molecular clouds on small scales.
They are an essential component of the early stages of star formation. As the collapsing protostellar system
expels an excess of angular momentum in the form of bipolar protostellar jets along the rotation axis of the system, 
the jets entrain and accelerate ambient cloud material to form the protostellar outflows. The outflows emit profusely in
molecular lines revealing to observers information about the local ambient cloud environment and the embedded driving
sources\,\citep{2013MNRAS.433.2226R}.
Although observations suggest outflows may have little effect on the largest scale turbulence in molecular clouds 
\citep[e.g.][]{2004RvMP...76..125M,2009ApJ...707L.153P,2012ApJ...744...32Z,2013MNRAS.433.2226R},
the compressive modes they generate may still be important at triggering star formation 
on small scales \citep{2011A&A...529A...1S}.

In a study by \citet{2012ApJ...761..156F}, the authors found that both Mach number and 
turbulent forcing are the main parameters in controlling the SFR. 
They establish that turbulent compressive type forcing can increase the SFR up to an order of magnitude compared 
to solenoidal type forcing, and a Mach number change from 5 to 50 can increase the SFR by a factor of up to five. 
The effects of magnetic field were revealed to play a more minor role by only reducing the SFR by a factor of two.
Although playing a minor role, recent studies of models and simulations have shown that 
the presence of magnetic fields are still required in order 
to reduce the SFR rate to the levels observed in molecular clouds \citep{2014prpl.conf...77P}.

NGC 1333 is an extremely rich star forming region. 
Part of the Perseus molecular cloud complex, it is classed as a reflection nebula, covering an area 
$7^{\circ} \times 3^{\circ}$ in the sky, and estimated through VLBI methods 
to be 235$\pm$18 parsec distant \citep{2008PASJ...60...37H}.
Its vicinity and diversity allow observers to study all phases of the star formation process
including, for example, protostellar collapse \citep{2006A&A...453..145B}, 
and the accretion disks and jets of a protobinary system \citep{2011PASJ...63.1281C}.
Mapping observations through CO ($J = 3-2$) and ($J=1-0$) have shown the structure of NGC 1333 to be dominated by 
many molecular outflows that have broken the cloud into low-density cavities separated by dense 
shells \citep{2000A&A...361..671K}.
A recent maser survey of NGC 1333 by \citet{2014AJ....148...80L} found several masers with no associated 
protostar or outflow detected suggesting they may indicate a very early evolutionary stage of star formation 
before the onset of the traditional outflow phase.
Can the energy deposited by the outflows in NGC 1333 maintain the observed level of turbulence?
\citet{2007ApJ...659.1394M} proposed that the observed level of turbulence in NGC 1333 
could be driven internally by outflows if certain conditions were met.
\citet{2008ApJ...674..336G} found evidence that Class II sources have dispersed surrounding 
molecular gas on short timescales suggesting outflows can ultimately destroy the NGC 1333 cloud.
However, \citet{2009ApJ...707L.153P} measured the turbulent energy spectrum of NGC 1333 $^{13}$CO maps and found 
slopes of $\sim -1.85$ with no flattening near the scale expected for outflow driving, suggesting turbulence in 
NGC 1333 may not be driven by outflows.
A JCMT survey of NGC 1333 by \citet{2010MNRAS.408.1516C} of C$^{18}$O data 
estimates that there is more energy introduced
by outflows than is observed in the bulk cloud, implying that it is not easy to feed high-energy outflow      
gas into turbulent motions.

In order to understand momentum transfer and observational power spectra in more detail, 
we can study it most readily and precisely through numerical simulations of turbulence.
In such simulations the turbulence is either driven globally
through Fourier space \citep[e.g.][]{2008ApJ...688L..79F,2012MNRAS.421.2531M,2013ApJ...763...51F},
or locally in real-space via protostellar outflows
\citep[e.g.][]{2007ApJ...662..395N,2009ApJ...695.1376C,2010ApJ...722..145C,2013MNRAS.432L..80M,2014ApJ...790..128F}.
In this paper we drive the turbulence in real-space with outflows using a parameter scheme from 
\citet{2007ApJ...659.1394M} based on the observational properties of NGC 1333.

Turbulence consists of eddies that transfer energy across a range of scales from the 
injection scale to the dissipative range whilst conserving momentum.
The degree of turbulence is dependent on the Reynolds number which 
describes the ratio of the inertial to viscous forces
\citep{2013ApJ...771..123B}. 
The turbulence phenomenon was first statistically derived by \citet{1941DoSSR..30..301K} 
through the assumption that the velocity field is a stochastic distribution with a constant energy 
transfer rate at all scales.
The result is the famous Kolmogorov energy spectrum for {\it incompressible} turbulence
\begin{equation}
 E(k) \propto \epsilon^{2/3} k^{-5/3}
\label{K}
\end{equation}
where the energy power spectrum, $E(k)$, is proportional to 
the energy input rate per mass $\epsilon$, and the wavenumber, $k$ (also 
known as the inverse length scale; $k = 2\pi/\lambda$), 
always possesses a negative power-law index of -5/3 \citep{2014NPGeo..21..587F}.

However, within the astrophysical regime 
-- such as molecular clouds -- turbulence is highly supersonic and compressible 
and the measured energy power spectra have been shown to clearly deviate from the incompressible -5/3 value. 
It was considered that the physics may be more appropriately represented by the Burgers turbulence model.
The Burgers model is related to a fully analytical description of 1D shocks in a 1948 paper by Burgers 
as an exact solution to the Navier Stokes equations.
The fundamental difference of the Burgers turbulence model to Kolmogorov turbulence is that 
energy is transferred from large scales to small scales via shocks rather than a cascade through eddies 
\citep{2003ApJ...590..858V}.
It envisages many shocks possessing step functions, and their Fourier transform is
proportional to $k^{-1}$ and the resulting velocity power spectrum is
\begin{equation}
 P_v(k) \propto k^{-2} .
\label{K}
\end{equation}

In recent years, much work has focused on finding a universal scaling law to represent supersonic 
compressible turbulence in the same way as the Kolmogorov law closely defines incompressible turbulence. 
A power spectrum based on the density is not suitable as simulations have shown a dependence of density with Mach number. 
For example, \citet{2005ApJ...630L..45K} performed isothermal Fourier driven turbulence simulations
driven by solenoidal modes for Mach numbers in the range of 1 to 12.
They found the density power spectral slope flattened with increasing Mach number due to the
development of sheet-like and filamentary density structures ranging from a slope of -1.73
for $\mathcal{M}_\textrm{RMS}$ = 1.2, to a slope of -0.52 for $\mathcal{M}_\textrm{RMS}$ = 12.
This shows that power index for density is a function of Mach number where at transonic velocities the 
density power spectrum may follow Kolmogorov turbulence.
Velocity power spectra may show a similar trend. 
Although modern supersonic compressible turbulence simulations typically possess velocity power-law exponents of -2 
as the turbulence would be shock dominated and representative of Burgers turbulence such as 
\citet{2013MNRAS.436.1245F}, 
early simulations of subsonic and transonic flows were fitted with spectral indices similar to
-5/3 \citep[e.g.][]{1994PhFl....6.2133P}. 

A highly supersonic velocity power spectral index is expected to be close to -2. 
However, a turbulent velocity domain is a combination of 
compressible (dilational) modes, and solenoidal (longitudinal or transverse) 
modes \citep[e.g.][]{2010A&A...512A..81F}.
Different driving mechanisms consist of different strengths and ratios of the constituent solenoidal and compressible
components.
It has been found to have a strong effect on the resulting density distribution with compressible driving leading to a broader
density distribution in comparison to one generated by purely solenoidal driving \citep{2008ApJ...688L..79F}.
This is because compressive forcing generates shocks leading to stronger compression and rarefaction   
in the computational domain \citep{2010A&A...512A..81F}.
It also leads to different spectral slopes.
For example, \citet{2010A&A...512A..81F} obtained a velocity power spectral slope of -1.86 for pure solenoidal driving 
and -1.94 for pure compressional driving.
Further study in \citet{2013MNRAS.436.1245F} showed it is difficult to directly compare the power spectrum of solenoidal 
driving to compressible driving due to different inertial ranges and different dependencies on numerical resolution. 
Due to the spectral differences of the underlying modes, a pure velocity spectrum could not be considered as a universal 
scaling relation.

An alternative relation is being studied as a potential universal scaling relation; 
the density weighted velocity power spectrum ($\rho^{1/3}v$). 
It was first tested by \citet{2007ApJ...665..416K} who found it to scale as $k^{-5/3}$, in-line with the Kolmogorov law.
But later simulations showed that there could still be a dependence on the driving mechanism. 
Fully compressible simulations by \citet{2010A&A...512A..81F} found $P(\rho^{1/3}v) \propto k^{-2.1}$.
Later, \citet{2011PhRvL.107m4501G} theoretically predicted that a strong compressible component 
would produce a $P(\rho^{1/3}v$) scaling relation with a slope of $-19/9$, which was in good agreement to the 
simulations.
The reason solenoidal driving would provide a different relation is explained by the previously 
mentioned analysis of \citet{2013MNRAS.436.1245F} regarding the inertial-range scaling of the 
compressible components.
In this paper we will examine the $\rho^{1/3}v$ scaling using our real-space outflow driven turbulence model. 

Unlike the majority of other numerical based turbulence studies which drive the turbulence on large global scales 
through Fourier-space \citep[e.g][]{2010A&A...512A..81F,2013MNRAS.436.3247K,2013MNRAS.436.1245F,2014ApJ...780...99Y},
we drive the turbulence exclusively in real-space on small local scales with spherical outflows.
Fourier-space driven turbulence has close ties to power spectra as the power spectra are themselves derived 
through Fourier space.
Our real-space model is not only more realistic physically, but also completely impartial 
to the power spectra analysis, and thus provides a useful comparison to results of Fourier-space driven turbulence.
Our aim is to quantify our real-space outflow model and see how it compares to Fourier driven turbulence models 
in terms of power spectra.

In reality, turbulence may be driven at more than one scale.
Simulations by \citet{2010ApJ...722..145C}
included an external isotropically driven turbulence in addition to their outflow driven turbulence model,
and obtained steeper velocity power spectra with a detectable knee in the power spectra
at the outflow driving scale.
Recently, \citet{2014ApJ...780...99Y} studied Fourier driven turbulence driven at two separate scales
simultaneously, effectively accounting for different driving mechanisms such as supernovae or galactic
rotation shear on larger
100 pc scales, and stellar winds and protostellar outflows on smaller parsec scales.
They found the effects of the smaller scale driving may not be easily observed through the kinetic
energy spectrum unless it is of comparable energy to the larger scale driving mechanism.
However, due to our localised driving, our method can only be used to
assume small scale local turbulent driving near a star-forming clump within a molecular cloud.

Our model produces an average density power spectrum slope of $\sim -1.2$, a lower value than 
is usually found in other works.
On the contrary, we measure the average velocity power spectrum slope to be -2.0, a value agreeable with that 
of Burgers shock dominated turbulence.
When we plot the density weighted velocity power spectrum ($\rho^{1/3}v$), we obtain a slope of -1.6.
This value is slightly shallower than the pure solenoidal driving case of \citet{2013MNRAS.436.1245F},
and significantly shallower than the  $k^{-19/9}$ value derived by \citet{2011PhRvL.107m4501G},
and obtained through simulations by \citet{2013MNRAS.436.1245F} 
for purely compressive driving. It suggests a fundamental difference between Fourier based driving and 
real-space driving. However, our measured slope of -1.6 may be comparable to the Kolmogorov scaling value 
of -5/3 ($\sim 1.67$). The small difference in the measurement may not be significant given the uncertainties from
time variations and systematics introduced by our relatively low resolution and resulting short inertial range 
for fitting the slope.

In terms of compressional and solenoidal components,
we find strong compressional modes located at active outflow bowshocks and instances of outflow-outflow collision.
Solenoidal modes are also generated at these locations, mostly through curved shocks, 
before dispersing throughout the computational domain.
Overall we find compressional modes have a slightly shallower power-law slope, 
suggesting these modes are more efficient at transferring energy from large scales to small scales. 
This is supported through the ratio of the solenoidal to compressional modes
\citep[See][]{2008ApJ...688L..79F}, 
which shows solenoidal modes are created through our outflow-outflow shock interactions and ultimately
slightly dominate on average throughout the computational domain in an attempt by the system to reach 
an equilibrium based on the degrees of freedom of the different modes.
However, at instantaneous times, both solenoidal and compressional components can be 
highly variable and interchangeable during the actively driven simulation.

The layout of our paper is as follows;
We describe our computational code and model setup in \S~\ref{method}, 
numerical results and power spectra are presented and discussed in \S~\ref{results}, 
and we conclude in \S~\ref{conclusion}.

\section{Numerical Methods} 
\label{method}

\subsection{Numerical code and basic equations}

The simulations presented here were performed using an isothermal multi-dimensional 
fixed-grid code based on the total variation diminishing scheme, known as the {\sc tvd} code.
It is an explicit finite-difference code and uses a second-order accurate Roe-type up-winded
Riemann solver to calculate the inter-cell flux\,\citep{1999ApJ...514..506K}.

\citet{2009A&A...508..541K} compared the performance of the {\sc tvd} code against 
three contemporary grid-based codes -- {\sc enzo, flash,} and {\sc zeus} -- through simulations of
decaying, isothermal, supersonic turbulence where each code used the same initial conditions.
It was found that the {\sc tvd} code produced comparable results to the others, 
and more efficiently with shorter run-times.
This efficiency makes the {\sc tvd} code particularly suitable for our outflow driven turbulence simulations. 
As there are a wide range of velocities on the grid at any given time, 
from the supersonic motion of the active outflows to the subsonic motions of decaying 
turbulence in remoter parts of the grid, the computational time-step is constrained by the fastest 
motions and the simulation must also be run for many dynamical timescales. Therefore use of a computationally 
efficient code is preferable.

We use an isothermal version of the {\sc tvd} code.
One may expect dynamical processes in molecular clouds to include molecular
cooling and chemistry.
However, a study of molecular turbulence by \,\citet{2006MNRAS.368..943P} found that the isothermal approximation
can be sufficient as the cooling timescale is shorter than the dynamical timescale. 

Previous works concerning numerical turbulence simulations have included magnetic fields
\citep[e.g.][]{2006ApJ...640L.187L,2010ApJ...709...27W,2011MNRAS.410L...8C,2014ApJ...780...99Y,2014ApJ...790..128F}. 
In this paper, we however only show spectral characteristics of a hydrodynamic turbulent flow driven by outflows 
in real-space. In a later study, we will include magnetic fields. 

The isothermal {\sc tvd} code solves the following two equations of hydrodynamics 
presented in Euler conservation form.
The conservation of mass,
\begin{equation}
 \partial_t \rho + \nabla \cdot \left( \rho \vec{v} \right) = 0
\end{equation}
and the conservation of momentum,
\begin{equation}
 \partial_t \vec{v} + \vec{v} \cdot \nabla \vec{v} = c_s^2 \frac{\nabla \rho}{\rho} + \vec{F}
\end{equation}
where $\rho$ is density, $t$ is time, $c_s$ is the isothermal sound speed, $\vec{v}$ is the flow velocity vector, 
and $\vec{F}$ is the outflow forcing term.

Our primary simulation is performed on a fixed uniform grid of 1024$^3$ cells with periodic boundary conditions.
This effectively emulates a small sub-set of a larger molecular cloud.
We repeated the simulation at lower resolutions of 512$^3$ and 256$^3$ cells for a convergence test 
as resolution is an important factor when analysing data via power spectra in 
order to resolve an inertial range sufficiently \citep[e.g.][]{2010A&A...512A..81F}. 

\subsection{Forcing scheme}

Our turbulence forcing model proceeds as follows; 
A 3 dimensional (3D) computational domain is initialised with a uniform gaseous medium.
Different points within the domain are randomly selected. At each randomly selected point, the mean
density, $\bar{\rho}$, of a volume defined by the outflow radius, $R$, is determined.
If the mean density of this volume is larger than the global mean, $\rho_0$, and no point within the
volume is below a lower-threshold value 0.001$\rho_0$, then it is chosen to be the location of an outflow.

The momentum, $P_j$, which an outflow should introduce is defined as follows,
\begin{equation}
 P_j = \iiint\limits_{<R} \rho (\vec{r} ) v_j (R) \left( \frac{r}{R} \right)^q \mathrm{d} V
\label{P}
\end{equation}
where $\rho (\vec{r})$ is the density within the outflow volume depending on the distance 
vector $\vec{r}$ from the
outflow center, and $v_j(R)$ is the outflow velocity at its outermost boundary, $R$.
The exponential $q$ parameter, set to $1.0$, attempts to mimic the observed `Hubble-law' type velocity
observed in protostellar outflows \citep[e.g.][]{2007prpl.conf..245A}.

As the total scalar momentum, $P$, which each outflow should introduce is known,
$v_j(R)$ can be expressed in terms of $P$ and the integral part of Eq.\,\ref{P}.
However, by setting a new velocity field, the total momentum of the outflow may not be conserved due to a
non-uniform density distribution existing within the chosen outflow radius, especially during later times of the simulation.
Therefore we calculate the added momentum and then subtract it within the radius $R$.
This ensures the vector sum of the added momentum is zero and the total momentum on the computational 
domain remains conserved.

\subsection{Dimensionless Parameter scheme}

We implement the same dimensionless parameter scheme first introduced 
by \citet{2007ApJ...659.1394M} and used in \citet{2013MNRAS.432L..80M} to provide a physical measure to our model.
This scheme was also used by \citet{2009ApJ...695.1376C,2010ApJ...722..145C} and ties our results to physical 
quantities. It is in fact based on the physical parameters measured from NGC 1333.
The dimensionless quantities of mass, $m_0$, length, $l_0$, and time, $t_0$, are defined as follows;
\begin{equation}
 m_0 = \frac{\rho^{\frac{4}{7}} P^{\frac{3}{7}} } { S^{\frac{3}{7}} }, \mbox{ } l_0 = \frac{P^{\frac{1}{7}} } { \rho^{\frac{1}{7}}  S^{\frac{1}{7}} }, \mbox{ and } t_0 = \frac{\rho^{\frac{3}{7}}}{ P^{\frac{3}{7}}  S^{\frac{4}{7}} } ,
\end{equation}
with parameters of density, outflow momentum, and outflow rate per unit volume set as  
$\rho$ = $2.51\times{10}^{-20}$ g\,cm$^{-3}$, $P$ = $3.98\times{10}^{39}$ g\,cm$^{-3}$\,s$^{-1}$, and 
$S$ = $6.31\times{10}^{-68}$ cm$^{-3}$\,s$^{-1}$, respectively.
The resulting physical unit quantities we use are thus;
$m_0$ = $18.7$ M$_{\odot}$, $l_0$ = $0.37$ pc, $t_0$ = $0.34$ Myr, and $v_0$ = ${l_0}$/${t_0}$ = $1.064$ km\,s$^{-1}$.
With these values the local sound speed is 0.587 km\,s$^{-1}$.
As in \citet{2013MNRAS.432L..80M} we set each side of the computational box as 
$L = 8\,l_0$ (3.96 pc), but this time we 
choose an outflow radius of $R$ = 0.35\,$l_0$ (a lower limit necessary for numerical stability of our code at 1024$^3$), 
and found it sufficient to run the simulation for time 2.0\,$t_0$\,($\sim$0.68 Myr). 

\subsection{Power Spectra}
\label{power}

In the next section we will analyse our turbulent data in terms of time-averaged compensated power spectra.
The power spectra are time averaged over a total of 50 output dumps between time 1.0 -- 2.0\,$t_0$,
a range where the turbulence has fully saturated whilst being actively driven.
The compensated power spectra method is often used
in turbulence studies\,\citep[e.g.][]{2009A&A...508..541K,2012ApJ...744...32Z,2013ApJ...763...51F},
as it aids in visualising the correct spectral index and inertial range.
The power spectrum of a simulated quantity, $P(k)$, is compensated by a factor, $k^{\delta}$, where
the exponent $\delta$ is 
chosen to negate the power-law slope, or in other words, the power-law exponent is the negative of $\delta$.
We only use the highest resolution simulation to determine the power-law slope. Due to the nature of our
real-space driving occurring on small scales, an appropriate inertial range is quite limited in comparison
to Fourier based driving.

The three-dimensional velocity field can hold additional information as to the nature of the velocity motions.
In real space a velocity field can be decomposed into three directions; $x$, $y$, and $z$. 
Correspondingly, in Fourier space, it can be decomposed into two; solenoidal (also known as `divergence free'),
and compressible (or `curl free').
The mathematical principle is based on the Helmholtz decomposition theory 
which states that a differentiable vector field with
divergence ($\nabla \cdot \vec{v}$) , and curl [$\nabla \times \vec{v}$], can be split
into its corresponding compressional and solenoidal components
\begin{equation}
 \vec{v}(r) = \vec{v}_s(r) + \vec{v}_c(r)
\end{equation}
where $ \nabla \times \vec{v}_c = 0$ and $\nabla \cdot \vec{v}_s = 0$.
The resulting solenoidal and compressional power spectra are defined as,
\begin{equation}
 P_s(k) = \frac{1}{N^3} \sum^{N} {|\vec{v}_s(k)|}^2
\end{equation}
\begin{equation}
 P_c(k) = \frac{1}{N^3} \sum^{N} {|\vec{v}_c(k)|}^2
\end{equation}
where $\vec{v}_s(k)$ and $\vec{v}_c(k)$ are the Fourier transforms of the solenoidal and compressional components, respectively.
Vorticity in the turbulent flow leads to solenoidal components, and converging flows generate
compressible components.
Studying these components and their ratios is a method now regularly used to characterise and
understand turbulence in numerical simulations \citep[e.g.][]{2008ApJ...688L..79F,2010A&A...512A..81F,2012MNRAS.421.2531M,2013MNRAS.436.1245F}.

\section{Results}
\label{results}

\subsection{RMS evolution and 2D visualisations}

\begin{figure}
  \epsfxsize=8cm
    \epsfbox[40 25 495 295]{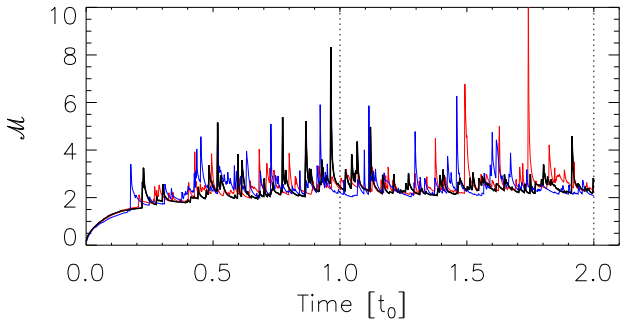}
\caption[]
{Root-mean-square Mach number evolution with time during our simulation at all resolutions;
1024$^3$ (black), 512$^3$ (blue), and 256$^3$ (red).
The turbulence saturates after a time of 0.5\,$t_{0}$.
The vertical lines at 1.0\,$t_{0}$ and 2.0\,$t_{0}$ indicate the time intervals between which we consider 
our time-averaged power spectra.
The average RMS Mach number between time 1.0--2.0\,$t_{0}$ is measured as 2.456 for the 1024$^3$ simulation, 
2.596 for the 512$^3$, and 2.460 for the 256$^3$}
\label{rmsplot}
\end{figure}

Our simulation quickly generates and maintains supersonic turbulence at a saturated level. This can be seen 
in the plot of root-mean-square (RMS) Mach number ($\mathcal{M}_\textrm{RMS}$) evolution with time in Figure\,\ref{rmsplot}. 
For comparison, the $\mathcal{M}_\textrm{RMS}$ evolution of the 512$^3$ and 256$^3$ simulations are plotted 
as blue and red lines, respectively.
In all cases supersonic turbulent motions become fully saturated after a time of 0.5\,$t_{0}$. 
We run the simulations to a time of 2.0\,$t_{0}$, at which point we have obtained 50 output dumps 
of fully saturated turbulence 
between a time of 1.0--2.0\,$t_{0}$ for use in the time averaged power spectra analysis to follow.
The high-velocity spikes are a consequence of our real-space driving model.
When an outflow is added to the grid, a small region may consist of very low-density cells.
Through the conservation of momentum, these cells would have a much higher velocity than average leading to a 
high velocity spike. However, the high velocity cells are highly localised and do not effect the global turbulence. 

\begin{figure}
  \epsfxsize=8.5cm
    \epsfbox[0 0 400 350]{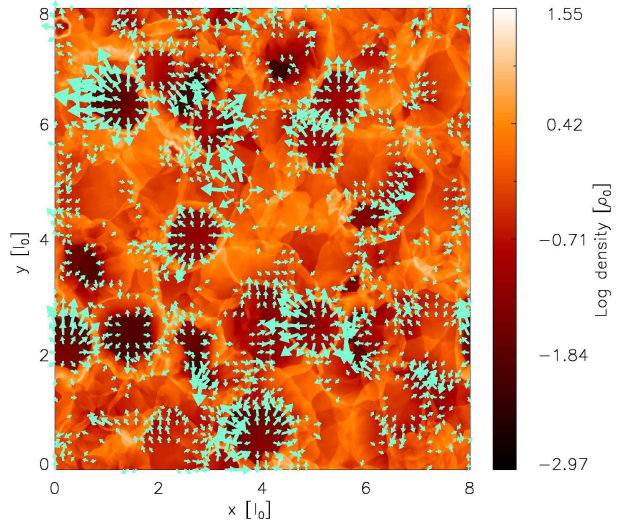}
\caption[]
{Slice through the 3D volume depicting log of density at time 2.0 $t_{0}$ for the 1024$^3$ simulation.
The outflows sweep the ambient gas into thin shells which interact with other shells creating the filamentary structure
interspersed with lower density voids. Arrows represent the velocity field vectors for high velocities.}
\label{denslice}
\end{figure}

\begin{figure}
  \epsfxsize=8.5cm
    \epsfbox[0 110 380 440]{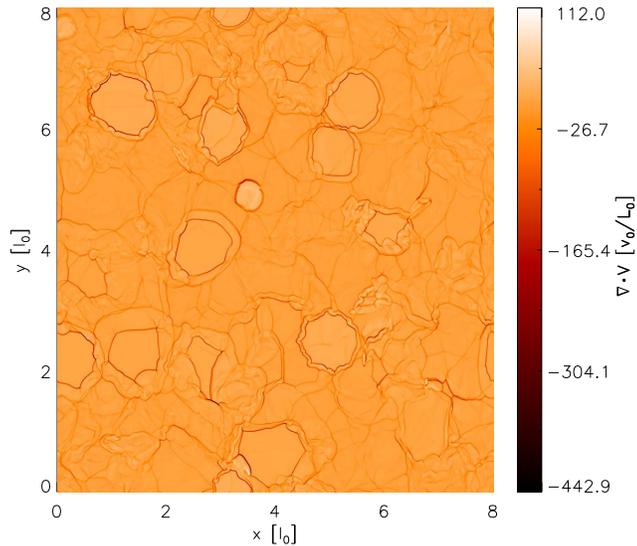}
\caption[]
{Slice through the 3D volume depicting real-space divergence ($\nabla \cdot \vec{v}$) data at time 2.0 $t_{0}$ for the 1024$^3$ simulation.
Expanding flows have positive divergence (light colour), whereas converging flows have negative divergence (dark colour).
This highlights the fact that the strongest shocks occur at the bow-shocks of expanding outflow cavities.}
\label{divslice}
\end{figure}

\begin{figure}
  \epsfxsize=8.5cm
    \epsfbox[0 110 380 440]{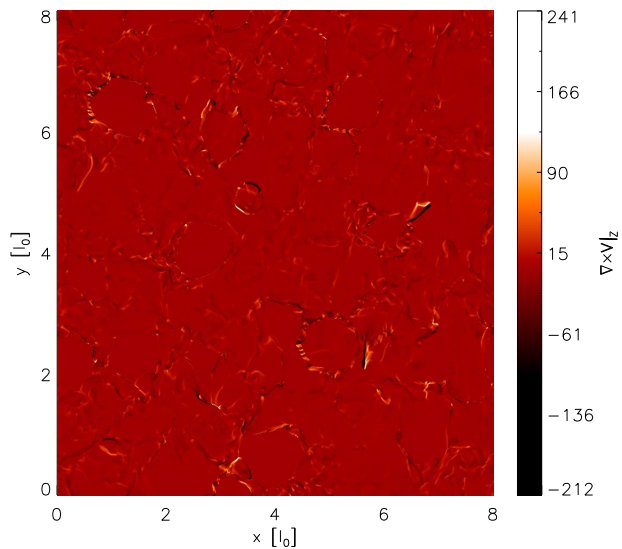}
\caption[]
{Slice through the 3D volume depicting the $z$-component of the curl, $\nabla \times \vec{v}$, 
at time 2.0 $t_{0}$ for the 1024$^3$ simulation.
This plot highlights the intricate structure of rotational motions which have been 
generated within the entire volume.}
\label{curlslice}
\end{figure}

To aid visualisation of the data, 
a slice through the 3D data cube representing the log of density at time 2.0\,$t_{0}$ is shown in Figure\,\ref{denslice}. 
Here we see the turbulent filamentary structure as well as newly expanding low-density outflow cavities. 
The expansion of the outflow cavities are highlighted by the arrows representing the velocity field on the 
displayed 2D plane for high velocities. 
These velocity vectors show the velocity profile of the outflows range from zero velocity at the interior, to 
maximum velocity at the boundary.
We see that high velocities are not sustained far from the outflow sources suggesting most of the momentum is lost 
through the bow-shock and converted into slowly expanding shells. 
This phenomenon is more clearly seen in Figure\,\ref{divslice}, which plots the divergence ($\nabla \cdot \vec{v}$) 
of the velocity field. 
The divergence field highlights expansion flows as positive values (light colour) 
and areas of compression as large negative values (dark colour). 
We see the strongest shocks occur around the boundaries of the new outflow cavities 
which are immersed in a filamentary structure of older 
weaker shocks. The weaker shock structures are reminiscent of the `fossil cavities' of \citet{2006ApJ...653..416C}.
In a complimentary fashion, we can plot the curl ($\nabla \times \vec{v}$) of the velocity field, where the 
curl represents rotational motions.
Figure\,\ref{curlslice} shows the $z$-component of the curl
$\nabla \times \vec{v} $ revealing 
that whilst most of the domain is saturated by small absolute values of rotational motion, 
interacting regions possess extreme values.

\subsection{Compensated power spectra}

\begin{figure}
  \epsfxsize=8cm
    \epsfbox[40 20 525 440]{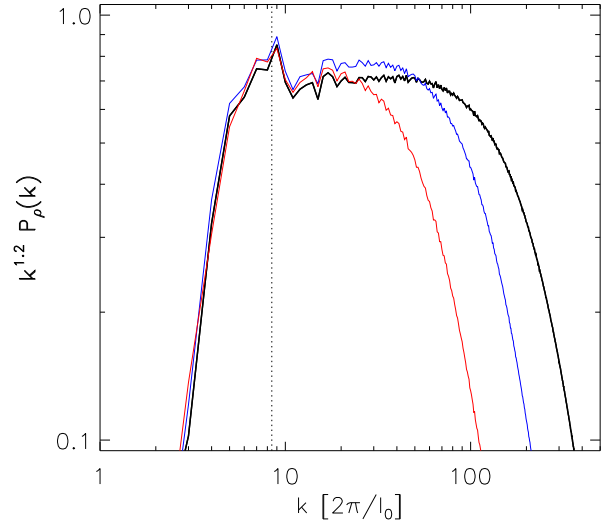}
\caption[]
{Compensated power spectrum of the time averaged density data for the 1024$^3$ simulation (black).
The normal power-spectrum, $\it{P}_{\rho}(k)$, is compensated by the factor $k^\delta$ where
$\delta$ = 1.2, represents the ideal power-law slope taken at an inertial range between $k=20-40$.
The corresponding compensated density power spectra for the 512$^3$ (blue) and 256$^3$ (red) simulations
are also shown at the same value of $k^\delta$.
A suitable inertial range is quite limited as our real-space driving method occurs on small scales only.
The vertical dotted line represents the outflow driving scale.
}
\label{compensatedden}
\end{figure}

\begin{figure}
  \epsfxsize=8cm
    \epsfbox[40 20 525 440]{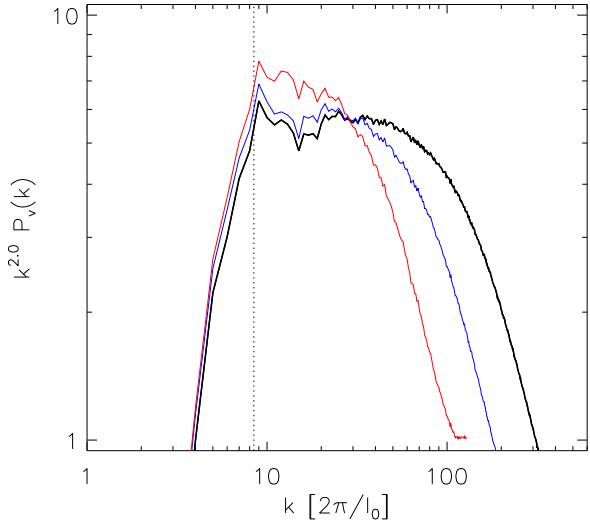}
\caption[]
{Compensated power spectra of the averaged time velocity data of the 1024$^3$ (black), 512$^3$ (blue),
and 256$^3$ (red) simulations.
In a similar fashion to Figure \ref{compensatedden}, the power-spectrum of velocity, $\it{P}_{v}(k)$,
is compensated by the factor $k^\delta$ where $\delta$ represents the ideal power-law slope between 
an inertial range of
$k=20-40$ for the 1024$^3$ data, which we find to be $\delta$ = 2.0.
At lower-resolutions, the velocity power spectra display poorer convergence compared to the density power spectra.
The vertical dotted line represents the outflow driving scale.
}
\label{compensatedvel}
\end{figure}

Now we analyse the full data cubes in terms of power spectra.
We plot the compensated power spectrum of 
density in Figure\,\ref{compensatedden}, and velocity in Figure\,\ref{compensatedvel}.

For our density power spectrum, $P_{\rho}(k)$, a slope of -1.2 was found to be the best fit to the data of 
the 1024$^3$ simulation 
between an inertial range of $k=20-40$ as seen in Figure\,\ref{compensatedden}.
The slope of the power spectrum flattens below this scale as usually seen in power spectra of outflow
driven turbulence simulations unless an additional larger scale driver is present \citep[e.g.][]{2014ApJ...780...99Y}.
The vertical dashed line represents the outflow driving wave-number, $k_{outflow}$/$k_0$. This is the scale 
on which the outflows introduce their energy.
It is defined as the outflow size expressed in wavenumber $k_{outflow}=2\pi/R$, 
divided by the size of the computational box 
expressed in wavenumber, $k_0=2\pi/L$.
We also plot the results of the lower resolution 512$^3$ (blue) and 256$^3$ (red) simulations. 
With decreasing resolution there is a decrease in the number of high-frequency wave-numbers, limiting the 
suitable inertial range and showing the importance of high-resolution simulations for 
accurate power spectral analysis.

\citet{2007ApJ...659.1394M} predicts the spectral slope of density would increase in regions 
where outflow momentum is captured more effectively.
The many low-density cavities created by the strong shocks of the outflows on our grid may 
translate to a lower than expected power spectrum. 
A large volume of low-density gas in real-space translates as an absence of lower frequency wave-numbers 
in Fourier space and hence a shallowing of the spectral slope. 

Our result is intermediate to that of \citet{2005ApJ...630L..45K} 
whose Fourier driven simulations 
obtained density spectral slopes of -1.73 for $\mathcal{M}_\textrm{RMS}$ = 1.2 turbulence, 
to a flatter -0.52 for $\mathcal{M}_\textrm{RMS}$ = 12.0. 
The authors accounted for this flattening as a consequence of the formation of dense filaments 
and sheets as Mach number increases.
If we assume a smooth relationship between spectral slope and Mach number in the simulations of 
\citet{2005ApJ...630L..45K}, we can perform interpolation to determine the spectral slope 
their model should provide if their turbulence was driven at the same Mach number as our model. 
After interpolation using a cubic polynomial fitted to their data points 
we determine that a $\mathcal{M}_\textrm{RMS}$ = 2.456 flow
would lead to a spectral slope of -1.3. 
This is slightly steeper than our value of -1.2 for the same Mach number. Several factors may account 
for this difference (e.g. driving mechanism and chosen inertial range).

The compensated power spectrum of velocity is presented in Figure\,\ref{compensatedvel}. 
Here we find the power spectrum slope for the 1024$^3$ simulation is best compensated by a $\delta$ exponent of 2.0 
for an inertial range of $k=20-40$. 
This time there is greater deviation between the spectral index at different resolutions. 
The lower-resolutions appear to show more power between the inertial ranges of $k=10-30$.
This is also noticeable in the Figure 6 of \citet{2013MNRAS.436.1245F} who analysed power spectra
of several resolutions up to an extremely high-resolution of 4096$^3$. However, our compensated spectral
slope is comparable within the chosen, albeit limited, inertial range of $k=20-40$ at our highest resolution 
as the 1024$^3$ resolution provides no suitable inertial range below $k=20$.
Overall the value of -2.0 for the slope suggests the velocity spectrum represents 
that of Burgers type turbulence and hence is dominated by strong shocks. 
This value of spectral index of velocity is frequently found in many previous studies at 
higher Mach numbers. It was obtained, for example, by \citet{2007ApJ...665..416K} with Mach number 6 driving, 
and \citet{2013MNRAS.436.1245F} with Mach number 17 driving. 
In comparison, our RMS Mach number is relatively low at 2.456. This suggests that 
the total velocity power spectrum is not dependent on Mach number once a turbulent flow is supersonic.

\subsection{Density weighted velocity power spectra}

Density weighted velocity power spectra ($\rho^{1/3}v$) have been proposed as a way for the Kolmogorov law to 
hold for compressible turbulence due to the mixed quantity statistics involved. 
A detailed study of density weighted velocity power spectra was recently
performed by \citet{2013MNRAS.436.1245F} and revealed it may not be a universal scaling relation as first thought.
Although it is expected that $P(\rho^{1/3}v) \propto k^{-5/3}$ for low resolutions,
the authors found a density weighted velocity power spectrum of $k^{-1.74}$ for pure solenoidal driving, 
and a steeper $k^{-2.10}$ for pure compressional driving. 
The reason suggested by the author for this difference is that compressible turbulence has a wider inertial range and
depends on the driving scale possibly through compressible 
shocks crossing multiple scales, unlike shocks from solenoidal driving. 
Therefore it is highly dependent on numerical resolution.

\begin{figure}
  \epsfxsize=8cm
    \epsfbox[20 20 520 440]{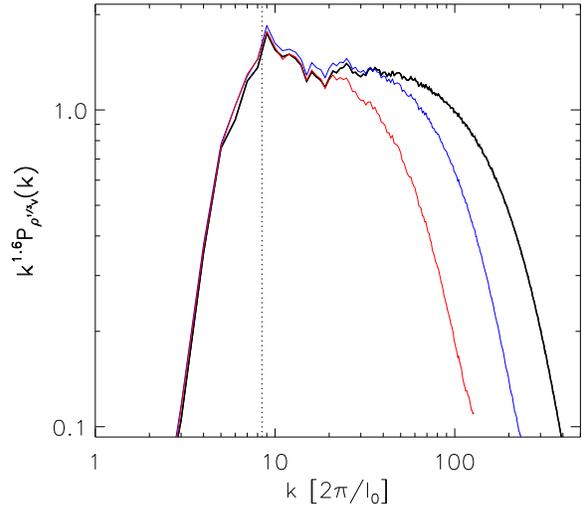}
\caption[]
{ Average density-weighted velocity power spectrum ($\rho^{1/3}v$) of the 1024$^3$ simulation.
We fit a slope of -1.6 between the inertial range of $k=20-40$ for the 1024$^3$ simulation.
The result of the 512$^3$ (blue) and 256$^3$ (red) simulations are also shown.
The vertical dotted line represents the outflow driving scale.
}
\label{rhovelspectra}
\end{figure}

When we plot the average density weighted velocity power spectrum of the 1024$^3$ simulation in 
Figure\,\ref{rhovelspectra}, we 
obtain a spectral slope of -1.6 in the previously chosen inertial range of $k=20-40$. 
We may have expected our result to be between the $k^{-1.74}$ pure solenoidal and  $k^{-2.10}$ pure 
compressional extremes of \citet{2013MNRAS.436.1245F} as our real-space driving is a combination of both 
solenoidal and compressional modes. 
But our result is actually shallower than their pure solenoidal limit.
In addition, our result is slightly shallower than the -5/3 ($\sim 1.67$) scaling as was initially 
expected by \citet{2007ApJ...665..416K},
although the difference may not be significant given the short inertial fit range and uncertainties 
arising from possible time variations and resolution effects of our model.
As our localised real-space driving model mostly agrees with Fourier based driving through the velocity field, it 
is likely the deviation in $P(\rho^{1/3}v)$ could be due to our density field where either our real-space 
localised driving may create a different density field morphology than one based on global Fourier driving, 
or the density field is very dependent on Mach number as 
was found in \citep{2005ApJ...630L..45K}.

\subsection{Solenoidal and compressional velocity components}

\begin{figure}
  \epsfxsize=8cm
    \epsfbox[20 20 520 440]{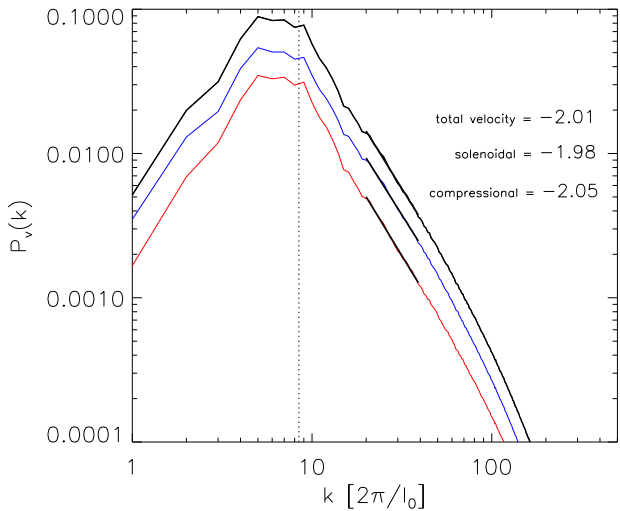}
\caption[]
{Averaged total velocity power spectrum (black line) and its 
solenoidal (blue) and compressional (red) components for the 1024$^3$ simulation.
Vertical dotted line represents the outflow driving scale.
We average velocity data of 50 output dumps between time 1.0--2.0\,$t_{0}$.
Power-spectra slopes measured over an inertial range of $k=20-40$.
We see a slight increase of power in the low wavenumbers above the outflow driving scale due to the 
physical expansion of the outflows. 
They effectively transfer power from their initial driving scale to a larger scale. 
This effect is not detectable through the compensated power-spectra plots as the compensated plots operate by 
visually enhancing the high wavenumbers at the expense of the low wavenumbers.
}
\label{PSvel1024}
\end{figure}

We can also compare our turbulence model to that of Fourier driven turbulence in more detail by 
separating the velocity field into 
the corresponding solenoidal and compressional components. 
The total velocity power spectrum is presented in Figure\,\ref{PSvel1024} in addition to its constituent solenoidal (red) 
and compressional (blue) components.
Once again, we see a slope of $\sim$-2.0 fits the total velocity power-spectrum between the inertial range of $k=20-40$. 
Our compressible modes slope of -2.05 is slightly steeper than the solenoidal modes slope of -1.98. 
This trend was also found in \citet{2007ApJ...665..416K}, who obtained a slope of 
-1.95 for the total velocity, -2.02 for the compressional components, and -1.92 for the solenoidal components.

\begin{figure}
  \epsfxsize=8cm
    \epsfbox[0 15 560 510]{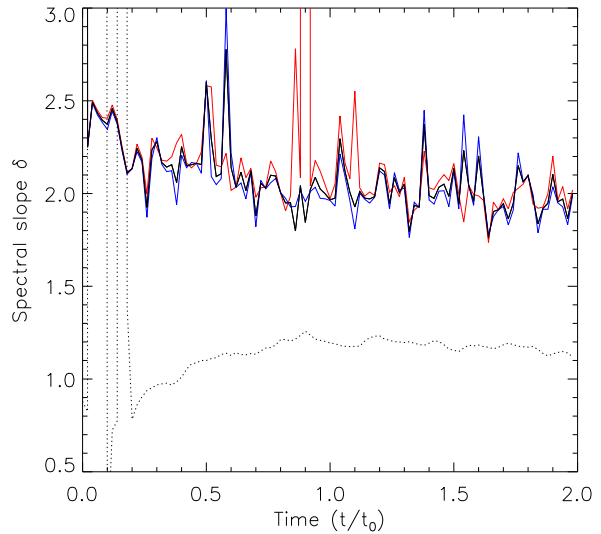}
\caption[]
{Slopes of the power spectra for the solenoidal (blue line) and compressional components (red line) with 
time for the 1024$^3$ simulation. 
The inertial range of $k=20-40$ as determined from the compensated velocity power spectrum 
was used here to plot the power spectral slope at each time interval.
The slopes of the density spectrum (dashed line) converge to an index value $\sim$ -1.2.
In comparison, the slopes of the velocity spectrum (solid black line) maintain large variations 
through the entire simulation, but have an average value of about -2.0 in the time range 1.0-2.0 
in agreement with Burgers type turbulence. 
}
\label{slopevstime}
\end{figure}

The small difference between the solenoidal and compressible spectra suggests that physically, 
compressional shocks may be slightly more efficient at dissipating energy compared to solenoidal shocks and 
hence the slightly steeper power-spectrum of the compressible modes.
This conclusion can be investigated in further detail by examining the interplay between the development of 
the compressional and solenoidal components with time. We do so in 
Figure\,\ref{slopevstime} where we plot the values of the spectral index between $k=20-40$ determined from 
the power spectra plots for both the density and velocity spectra with time over the entire simulation run-time.
After some initial spurious values at early times before the turbulence saturates, 
the index of the density power-spectrum (dotted line) smoothly builds up to maintain a 
relatively stable average value of $\sim -1.2$ after time 1.0\,$t_0$ until the end of the simulation.
This is in contrast to \citet{2007ApJ...665..416K} who found their density power-spectrum 
exhibited strong variations on short timescales.
However, in comparison our time evolution of the velocity power spectrum index is highly variable 
throughout the simulation. The solenoidal (blue) and compressional (red) component indexes are seen to sometimes vary 
by a large amount, yet the average spectral index fluctuates around a value of -2.0 during the saturated phase.
This suggests that the nature of our real-space actively driven turbulence is highly variable in terms of velocity, 
yet retains an average value of -2 representing Burgers strong shock type turbulence. 
Our spectral slope steepening is due new outflow events where we see a strong event can easily 
change the spectral slope.
Therefore, when determining power spectra of velocity fields with locally forced turbulent simulations, 
one should ensure to average over time otherwise the system at an 
instantaneous time may provide misleading results. 
This is due to the fact that 
the spectral index of velocity power spectra measured in observations could be changed a lot by localised strong outflows, 
supernova explosions, or any other energetic point-source driving sources.

\subsection{Ratio of the solenoidal to compressional velocity components}

Another method of quantifying the nature of our velocity field is measuring 
the relative strengths between the solenoidal to compressional components, 
as originally defined in \citet{2008ApJ...688L..79F} and \citet{2010A&A...512A..81F} as 
\begin{equation}
\mathcal{P}^\zeta_{ij} = \zeta\mathcal{P}^{\perp}_{ij} + (1-\zeta)\mathcal{P}^{\parallel}_{ij}  = \zeta \delta_{ij} + (1 - 2 \zeta) \frac{k_i k_j}{|k|^2}
\end{equation}
where a Helmholtz decomposition of the projection operator, $\mathcal{P}^\zeta_{ij}$, can represent a 
purely solenoidal field if $\zeta = 1$, or a purely compressible field if $\zeta = 0$, 
where the projection operators for the fully solenoidal and fully compressive modes are 
$\mathcal{P}^\perp_{ij} = \delta_{ij} - k_ik_j/k^2$, and $\mathcal{P}^\parallel_{ij} = k_ik_j/k^2$, respectively, 
with $\delta_{ij}$ representing the Kronecker delta.

\begin{figure}
  \epsfxsize=8cm
    \epsfbox[60 90 520 365]{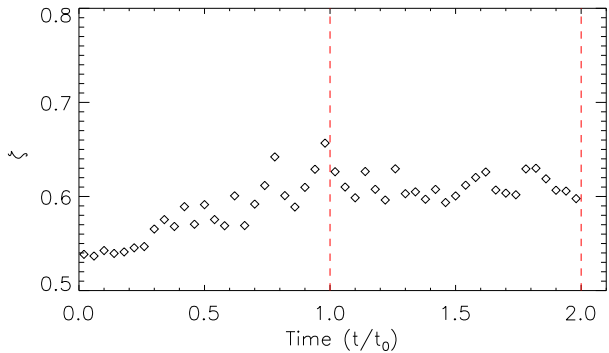}
\caption[]
{Evolution of the relative strengths of the compressional to solenoidal components, referred to as the $\zeta$ value,
during the 1024$^3$ simulation. We obtain an average $\zeta$ of 0.614 between a
time of 1.0 to 2.0 $t_0$ indicating that solenoidal modes slightly dominate throughout the simulation. It is
close to the expected 2/3 value based on the degree of freedom interpretation.
}
\label{xi}
\end{figure}

Although the authors defined it with respect to their forcing function, 
we define it based on the resulting turbulent velocity field at each output dump during the simulation.
A value of $\zeta$=0.0 would represent purely compressible turbulence, and a value of $\zeta$=1.0 would represent 
purely solenoidal turbulence.
We plot our $\zeta$ evolution in Figure\,\ref{xi}.
Just after time 0\,$t_0$ when a few outflows have been introduced to the grid but have not yet interacted, 
the $\zeta$ value is 0.535 showing each outflow in our model introduces almost equal amounts 
of compressional and solenoidal modes, although slightly skewed towards solenoidal modes.
After the outflows have interacted and the turbulence has become fully developed, we measure the 
average $\zeta$ to be 0.614 between time 1.0--2.0\,$t_0$. This value indicates 
solenoidal modes have increased and slightly dominate in our fully converged saturated turbulent velocity field.
Interestingly it is close to the expected value of 2/3 when we consider the degrees of freedom approach 
where statistically the system will try to maintain an equilibrium of 2/3 solenoidal and 1/3 compressional modes.
This arises because the compressible mode is longitudinal where there is only one degree of freedom (propagation direction).  
Whereas the solenoidal mode is transverse having two degrees of freedom 
(two perpendicular directions with respect to the propagation direction).
The slight discrepancy may be due to the fact that a constant strong compressional driving is biasing the natural balance.

\section{Conclusions}
\label{conclusion}

In this paper we study the power spectra of outflow driven turbulence through 
high-resolution isothermal simulations using the {\sc tvd} code.
Unlike the alternative method of Fourier-space driven turbulence that drives the 
turbulence on global scales, in our model the turbulence is driven 
locally in real-space by a series of spherical outflows. 
We use the parameter scheme of \citet{2007ApJ...659.1394M} in order to approximate the 
physical values of a typical star-formation cloud such as NGC 1333.
The resulting turbulence saturates with an average RMS Mach number of $\sim$2.5, and the 
time-averaged power spectrum as well as the relative strengths of the solenoidal to compressional 
components are investigated. 

An index of -1.2 is measured for our time-averaged density power spectrum. 
It is a lower value of density power spectral slope than found in subsonic compressible flows.
An explanation is the fact that strong shocks gather the gas together into a network of thin 
dense sheets and filaments. Then, in Fourier space, this translates to more power within the higher-frequency wavenumbers. 
Conversely the large volumes of low-density cavities lead to a depletion of lower-frequency wavenumbers. 
Both effects lead to a shallower power-law slope when viewing in Fourier space \citep{2005ApJ...630L..45K}.

In contrast, our total velocity spectral index is measured as $\sim$2.0, comparable to Burgers turbulence. 
This value is consistent with the recent work of \citet{2013MNRAS.436.1245F} 
who performed higher Mach number ($\mathcal{M}=17$) Fourier driven turbulence simulations with different 
total power.
This suggests the ability to obtain Burgers shock dominated turbulence through velocity spectra is quite robust 
despite variables of Mach number, resolution, and power.

The nature of power spectra and the Fourier transform allow us to readily separate the velocity field into 
compressional (colliding shocks), and solenoidal (rotational motions) components. 
We see that on average, the slope of the compressional power spectrum (-2.05) is slightly steeper than 
that of the solenoidal spectrum (-1.98). 
This can be explained by a combination of strong curved compressible shocks generating solenoidal modes, 
and compressible modes decaying more rapidly than
solenoidal modes at the small dissipation scales leading to a surplus of solenoidal power at small scales, and hence
a steeper compressible power spectral slope due to the resulting lack of compressible high-frequency wavenumbers.
The same result can be seen in the 
plot of relative strengths between the solenoidal to compressional components in Figure \ref{xi} where 
we obtain an average $\zeta$ value of 0.614, showing solenoidal modes slightly dominate after 
each outflow introduces almost an equal amount of solenoidal to compressional mode components at early times.
The Fourier space driven turbulence models of \citet{2013MNRAS.436.1245F} also show a steeper 
compressional than solenoidal spectrum.
Statistically, compressional modes possess 1 degree of freedom, whereas solenoidal modes 
possess 2 degrees of freedom. The system tries to maintain equilibrium of 1/3 compressible and 
2/3 solenoidal. This is consistent with our results.

The average density weighted velocity power spectrum ($\rho^{1/3}v$) of our model is measured with a spectral index of -1.6.
This value is slightly smaller than the Kolmogorov -5/3 scaling found in previous studies at similar resolution.
If uncertainties of resolution and the short inertial range fitting could be excluded,
it may be due to the nature of our real-space driving mechanism and the low value of the spectral index 
of the density field.
This may suggest power spectra reveal a fundamental difference between Fourier based driven turbulence 
and real-space driven turbulence. Fourier based driving could be considered as large scale driving; 
it stirs the entire computational domain equally. Then the turbulence must decay 
transferring energy from large scale to small scale.
In contrast real-space driving only stirs the computational domain in localised regions. Energy needs to be transferred 
up from small scale to large scale, against the natural direction of turbulence decay resulting in more energy remaining 
locked at small scales.
As the spherical outflows in our model expand, they transfer power to a larger scale than their initial injection scale. 
This transfer of power from a small scale to a larger scale is not due to turbulence, 
but the nature of our real-space driving model.

We tentatively propose that the density weighted velocity power spectrum ($\rho^{1/3}v$) is not a universal power 
spectrum. 
Once Mach number reaches a certain value, the index of total velocity is -2.0, even if driven with much greater Mach number. 
In contrast, the index of density power is variable. 
So a combination of density and velocity power would still be variable with Mach number.
We are currently investigating how the spectrum of $\rho^{1/3}v$ depends on the Mach numbers for a future publication.

Returning to the consideration of NGC 1333, our simulation results suggest that in an outflow-driven environment 
the solenoidal velocity component may be stronger than the compressible velocity component,
and the spectral slope of density power spectrum may be closely related to the global Mach number of the medium.

\section{Acknowledgements}

Numerical simulations were performed using a high performance computing
cluster in the Korea Astronomy and Space Science Institute (KASI).
S.-J.Y. acknowledges support by
Mid-career Research Program (No. 2012R1A2A2A01043870) through the
National Research Foundation (NRF) of Korea,
the DRC program of Korea Research Council of Fundamental Science and Technology,
the NRF of Korea to the Center for Galaxy Evolution Research (No. 2013-8-1583),
and the Yonsei University Future-leading Research Initiative of 2014-2015.
We thank the anonymous referee for useful comments.

\bibliography{tvd}

\label{lastpage}

\end{document}